\documentclass[final]{aipproc}
\usepackage{amsmath,amssymb,graphicx,bm}
\layoutstyle{6x9}

\newcommand{\lm}{\Lambda}
\newcommand{\lb}{\Lambda_{\rm b}}
\newcommand{\be}{\begin{equation}}
\newcommand{\ee}{\end{equation}}
\newcommand{\vnn}{V_{\rm NN}}
\newcommand{\vlowk}{V_{{\rm low}\,k}}
\newcommand{\fm}{\, \text{fm}}
\newcommand{\fmi}{\, \text{fm}^{-1}}
\newcommand{\mev}{\, \text{MeV}}

\newcommand{\la}{\langle}
\newcommand{\ra}{\rangle}

\begin{document}

\title{Three-nucleon interactions:\\
A frontier in nuclear structure}


\author{A.~Schwenk$^*$ and J.D.~Holt}{
address={TRIUMF, 4004 Wesbrook Mall, Vancouver, BC, Canada, V6T 2A3\\
$^*$E-mail: schwenk@triumf.ca}}

\begin{abstract}
Three-nucleon interactions are a frontier in understanding  
and predicting the structure of strongly-interacting matter in
laboratory nuclei and in the cosmos. We present results and discuss 
the status of first calculations with microscopic three-nucleon 
interactions beyond light nuclei. This coherent effort is possible 
due to advances based on effective field theory and renormalization 
group methods in nuclear physics.
\end{abstract}

\maketitle

\begin{figure}[t]
\includegraphics[scale=0.22,clip=]{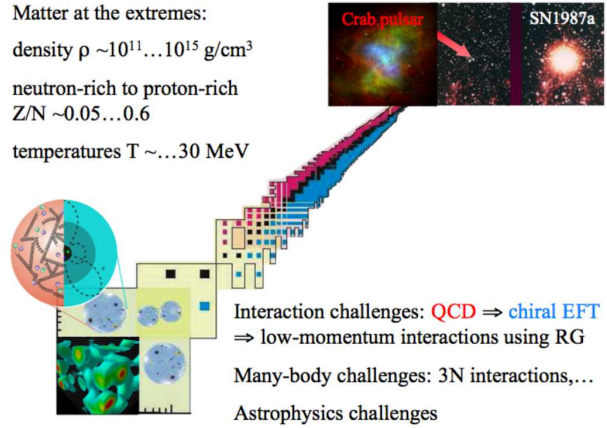}
\caption{From the QCD vacuum to laboratory nuclei to neutron stars and
supernovae.\label{strong}\vspace*{-4mm}}
\end{figure}

The physics of strong interactions extends over extremes in density,
neutron-to-proton imbalance and temperature (see Fig.~\ref{strong}).
Understanding and predicting the properties of these fascinating
forms of matter requires progress 
on fundamental problems in the theory of nuclear forces and in many-body
physics. In this talk, we highlight the key role of three-nucleon (3N) 
interactions for nuclear structure.

Nuclear interactions depend on a resolution scale, which we denote
by a generic momentum cutoff $\lm$, and the Hamiltonian is always
given by an effective theory for nucleon-nucleon (NN) and corresponding
many-nucleon interactions (with corresponding effective 
operators)~\cite{pionless,chiral,Vlowk}:
\be
H(\lm) = T + \vnn(\lm) + V_{\rm 3N}(\lm) + V_{\rm 4N}(\lm) + \ldots \,.
\label{Hamiltonian}
\ee
\noindent
This scale dependence is similar to the scale dependence of parton 
distribution functions, and shows that the effect of 3N interactions 
depends on this scale and the theoretical convention that enters all
parts of the Hamiltonian.

At very low momenta $Q \ll m_\pi \approx 140 \mev$, the details of pion
exchanges are not resolved and nuclear forces can be systematically
expanded in contact interactions and their derivatives~\cite{pionless}.
The corresponding pionless effective field theory (EFT) is extremely 
successful for capturing universal large scattering-length physics (with
improvements by including effective range and higher-order terms) in
loosely-bound or halo nuclei, reactions at astrophysical energies and
in dilute neutron matter (see e.g., Ref.~\cite{dEFT}).

For most nuclei, the typical Fermi momenta are $Q \sim m_\pi$ and 
therefore pion exchanges have to be included explicitly. In chiral
EFT, nuclear interactions are then organized in an expansion in 
powers of $Q/\lb$, where $\lb$ denotes the breakdown scale,
roughly $\lb \sim m_\rho$~\cite{pionless,chiral}.
The great advantage is that up to next-to-next-to-next-to-leading 
order (N$^3$LO) only two new 3N couplings enter, since parts
of the 3N force can be consistently determined by $\pi$N and NN 
couplings. The two 3N couplings, as well as the short-ranged NN 
couplings, depend on the resolution scale $\lm$ and for each $\lm$ 
can be fit to data.

\begin{figure}[t]
\parbox[c]{0pt}{%
\includegraphics[scale=0.56,clip=]{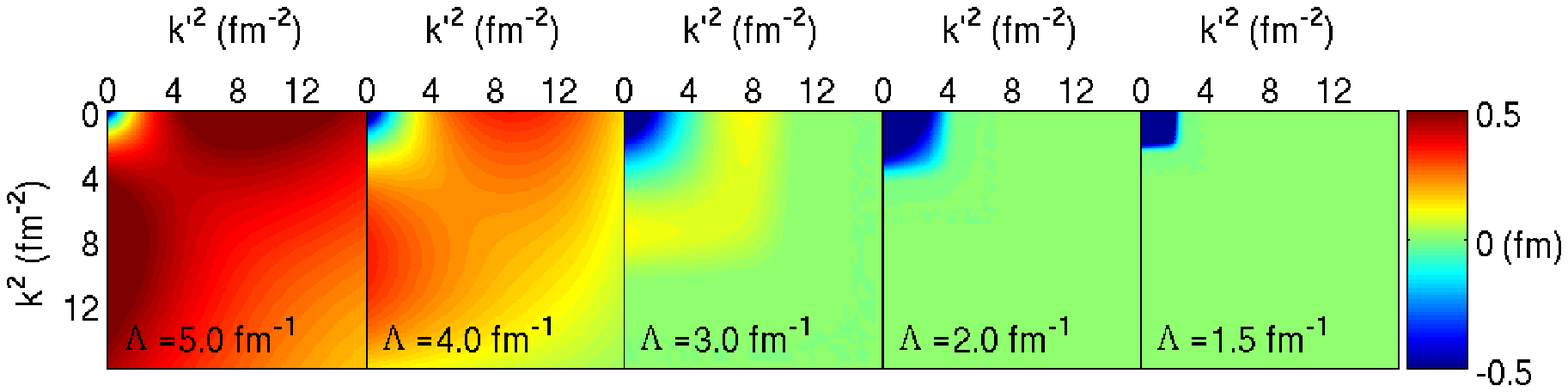}}
\hspace*{-6pt}
\raisebox{-110pt}{%
\includegraphics[scale=0.56,clip=]{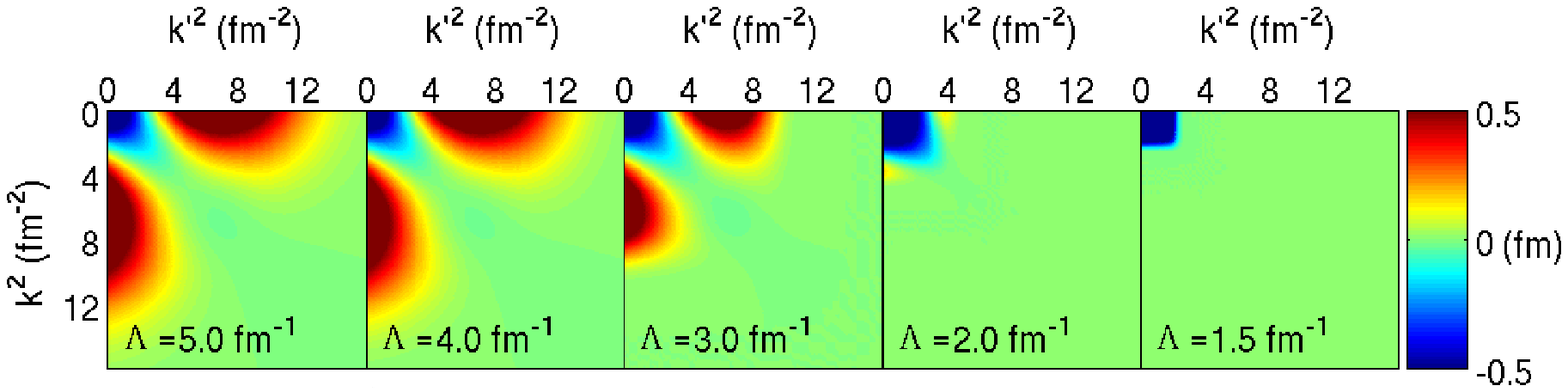}}
\caption{Evolution of the $^3$S$_1$ partial wave with a smooth $n_{\rm exp} 
= 8$ regulator at cutoffs $\lm = 5.0, 4.0, 3.0, 2.0$ and $1.5 \fmi$
(for details see Ref.~\cite{smooth}). The initial potentials are Argonne 
v$_{18}$~\cite{AV18} (top) and a 
N$^3$LO chiral EFT potential from Ref.~\cite{EM}
(bottom). The color scale ranges from $-0.5$ to $0.5 \fm$.\label{vlowk_3S1}
\vspace*{-6mm}}
\end{figure}

Using the renormalization group (RG), we can evolve an initial potential
to lower resolution by integrating out high momenta through discretized RG
flow equations or equivalent Lee-Suzuki transformations~\cite{Vlowk,VlowkRG}.
In the last two years, these methods have been refined to employ
smooth regulators~\cite{smooth} and similarity renormalization group 
(SRG) transformations~\cite{SRG1,SRG2}, which both have technical advantages 
in oscillator spaces.

\begin{figure}[t]
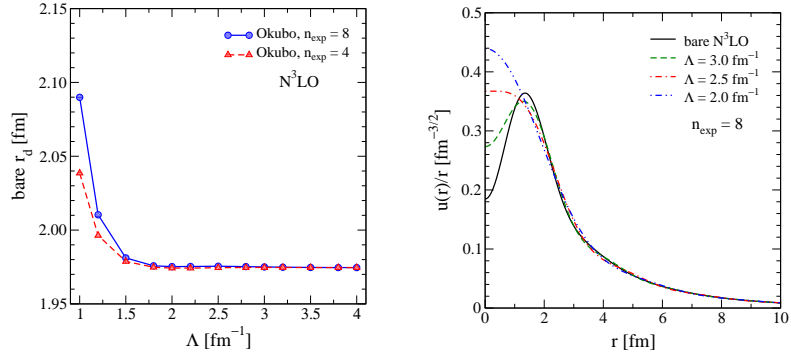

\includegraphics[scale=0.26,clip=]{deuteron_rsq}
\hspace*{6mm}
\includegraphics[scale=0.26,clip=]{deuteron_wf}
\caption{The deuteron rms radius $r_d$ calculated with the bare operator
as a function of the cutoff (left). The radius is only weakly dependent on
the cutoff for $\lm > 1.5 \fmi$, although the intrinsic resolution in the
S-state deuteron wave function changes substantially with $\lm$ (right).
For details see Ref.~\cite{smooth}.\label{deuteron}\vspace*{-6mm}}
\end{figure}

Changing the cutoff leaves observables unchanged by construction, but 
shifts contributions between the potential and the sums over intermediate
states in loop integrals. Since the sums are restricted by the intrinsic
resolution $\lm$, these shifts can weaken or largely eliminate sources of 
nonperturbative behavior such as strong short-range repulsion and short-range
tensor forces~\cite{nucmatt,Born}. 
As shown in Fig.~\ref{vlowk_3S1}, the evolution leads
to low-momentum interactions, generally known as ``$\vlowk$'', which become
universal for $\lm \lesssim 2 \fmi$ and have weak off-diagonal coupling.
Since the RG preserves the long-range parts of nuclear 
interactions~\cite{smooth}, this
leads to a weak renormalization of long-range operators, for example for 
the deuteron rms radius in Fig.~\ref{deuteron}.

\begin{figure}[t]
\includegraphics[scale=0.28,clip=]{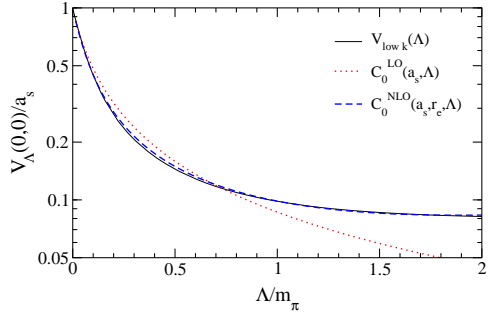}
\caption{Flow of $\vlowk(k'=0,k=0;\lm)$ compared to the corresponding 
momentum-independent contact interaction $C_0(\lm)$ at LO and NLO,
where this coupling is determined entirely from RG invariance and 
fits to the scattering length $a_{\rm s}$ (at LO) plus effective range
$r_{\rm e}$ (at NLO).\label{1s0lambda}\vspace*{-6mm}}
\end{figure}

\begin{figure}[t]
\includegraphics[scale=0.32,clip=]{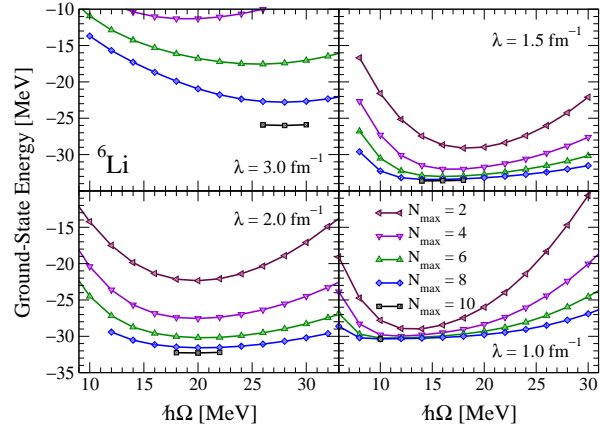}
\caption{Ground-state energy of $^6$Li as a function of the oscillator
parameter $\hbar\Omega$ at four different values of SRG resolution 
$\lambda = 3, 2, 1.5$ and $1 \fmi$. The NCSM results clearly show improved 
convergence with the maximum number of oscillator quanta $N_{\rm max}$ for
lower cutoffs. For details see Ref.~\cite{NCSM}.\label{Li6}
\vspace*{-5mm}}
\end{figure}

The evolution of chiral EFT interactions to lower cutoffs is beneficial 
for two reasons. First, the RG generates all higher-order short-range
contact interactions so that observables are exactly reproduced and
the theoretical uncertainty remains at the level of the truncation error
in the initial potential. We illustrate this in Fig.~\ref{1s0lambda}
by comparing the flow of $\vlowk(k'=0,k=0;\lm)$ with the corresponding 
momentum-independent contact interaction $C_0(\lm)$ in subsequent orders 
of pionless EFT.
Second, lower resolutions can be represented efficiently in oscillator
bases, and therefore lead to direct convergence in nuclear structure 
calculations~\cite{smooth}. This is demonstrated by the very promising
convergence for $N_{\rm max} \sim 10$ in NCSM calculations with SRG
interactions~\cite{NCSM} shown in Fig.~\ref{Li6}.

\begin{figure}[t]
\includegraphics[scale=0.24,clip=]{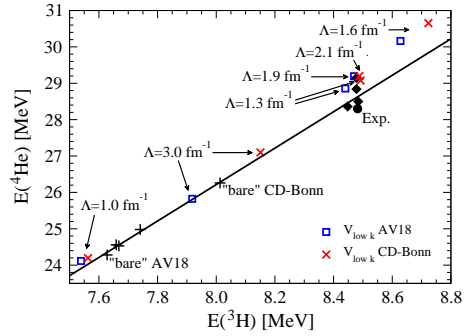}
\caption{Correlation of the $^3$H and $^4$He binding energies. The cutoff
dependence of the exact NN-only results with $\vlowk(\lm)$ explains
the empirical (solid) Tjon line. For details see 
Ref.~\cite{Vlowk3N}.\label{Tjon}\vspace*{-6mm}}
\end{figure}

\begin{figure}[ht!]
\includegraphics[scale=0.26,clip=]{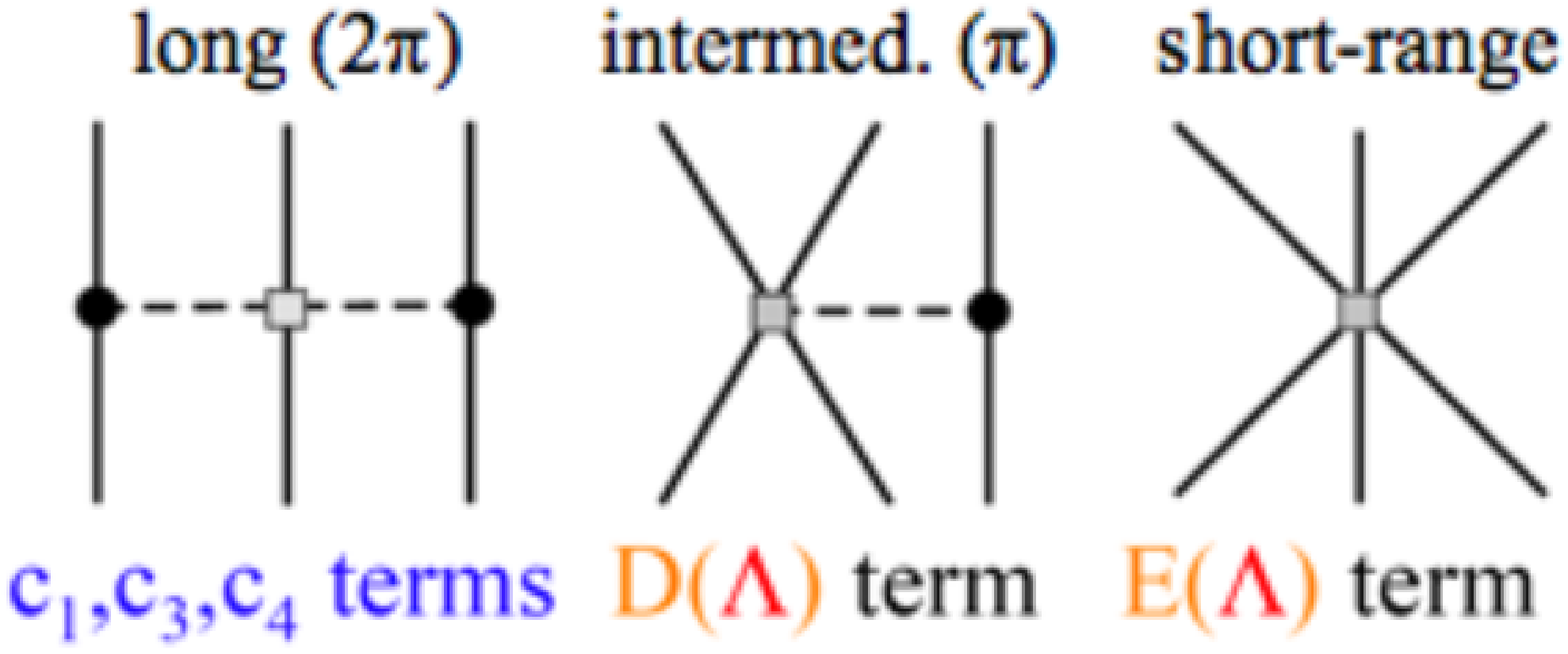}
\caption{Leading (N$^2$LO) 3N interaction in chiral EFT without explicit 
Deltas.\label{3N}\vspace*{-6mm}}
\end{figure}

$\vlowk(\lm)$ defines a class of NN interactions with cutoff-independent
low-energy NN observables. Consequently, any cutoff variation of observables
estimates the truncation errors due to neglected many-body interactions 
in $H(\lm)$ of Eq.~(\ref{Hamiltonian}). Figure~\ref{Tjon} shows that this
cutoff dependence explains the empirical Tjon line, that 3N interactions
are required by renormalization and needed to break off the line to
accurately describe the experimental $^3$H and $^4$He binding 
energies~\cite{Vlowk3N}. NN-only results also lead to Tjon lines in
medium-mass nuclei, where results truncated in 
oscillator shells as well as for different $\hbar \Omega$ lie approximately
on the same lines~\cite{Oxygen_Tjon}.

Three-nucleon interactions are crucial for many-nucleon systems. In
addition to the impact on binding energies, they play a central role
for spin-orbit and spin dependences, for isospin dependences of
neutron- and proton-rich systems, and they drive the density dependence
of nucleonic matter. The latter are pivotal for extrapolations to
the extremes of astrophysics. Three-nucleon interactions are also
a frontier in few-body scattering (for a critical discussion, see 
e.g., Ref.~\cite{Meyer}). Since 3N contributions are amplified
in nuclei, it may be necessary for controlled predictions to 
constrain 3N couplings with few- and many-body data. Therefore a
coherent 3N effort is needed with theoretical uncertainties.

\begin{figure}[t]
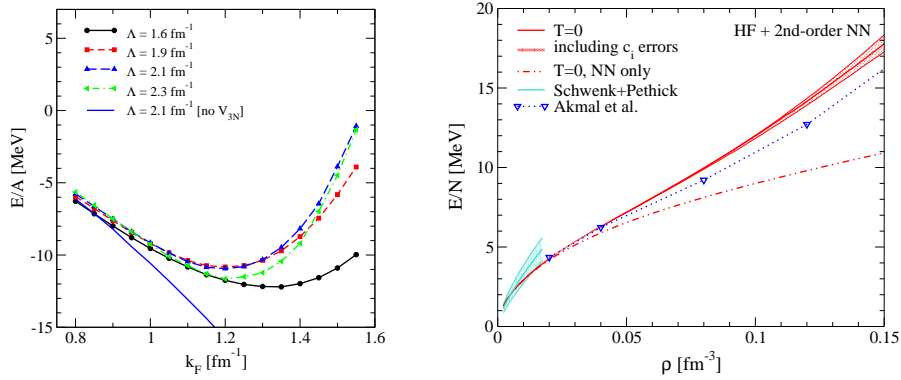

\includegraphics[scale=0.28,clip=]{paper_2ndorder_mstar_2+3_fullP_rev2}
\hspace*{6mm}
\includegraphics[scale=0.28,clip=]{energy_T0_v3}
\caption{Nuclear (left) and neutron matter (right) from low-momentum NN and 3N 
interactions based on Hartree-Fock plus dominant second-order 
contributions for various cutoffs. For details 
see Ref.~\cite{nucmatt,neutmatt}.\label{matter}\vspace*{-6mm}}
\end{figure}

In chiral EFT without explicit Deltas, 3N interactions start at N$^2$LO
or order $(Q/\lb)^3$~\cite{pionless,chiral} and their contributions are given
diagrammatically in Fig.~\ref{3N}. The low-energy constants for the
long-range parts relate $\pi$N, NN and 3N interactions, and the
determination from $\pi$N scattering is, within errors, consistent
with the extraction from peripheral NN waves. The present constraints
are $c_1 = -0.9^{+0.2}_{-0.5}$, $c_3 = -4.7^{+1.2}_{-1.0}$ and $c_4 = 
3.5^{+0.5}_{-0.2}$ (all in ${\rm GeV}^{-1}$)~\cite{Ulf}. 
In particular, $c_3$ and $c_4$ are
important for nuclear structure and have large uncertainties (see
Fig.~\ref{matter} for the impact on nucleonic matter).

\begin{figure}[t]
\includegraphics[scale=0.3,clip=]{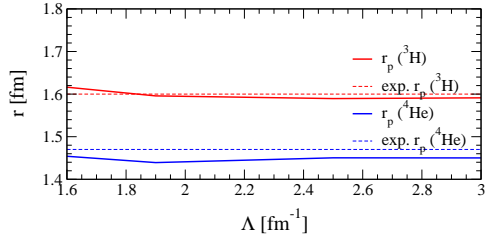}
\caption{The $^3$H and $^4$He radii are approx.~cutoff independent with
NN and 3N interactions~\cite{Andreas}.\label{radii}\vspace{-5mm}}
\end{figure}

\begin{figure}[t]
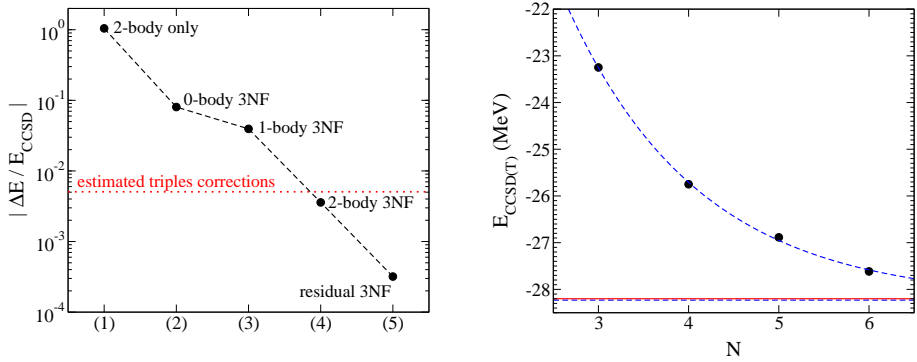

\includegraphics[scale=0.234,clip=]{v3_fig_new}
\hspace*{6mm}
\raisebox{-9pt}{%
\includegraphics[scale=0.234,clip=]{ccsdt_min_new}}
\caption{Relative contributions $|\Delta E / E|$ to the binding energy 
of $^4$He at the CCSD level from $\vlowk$ as well as normal-ordered
0-, 1-, 2-body and residual 3-body parts of the 3N interaction (left).
Convergence of the corresponding CCSD(T) results with the
number of oscillator shells (right). The extrapolated binding energy
$-28.23 \mev$ agrees well with the exact result $-28.20(5) \mev$. For
details see Ref.~\cite{VlowkCC}.\label{CC}\vspace*{-6mm}}
\end{figure}

\begin{figure}[t]
\includegraphics[scale=0.28,clip=]{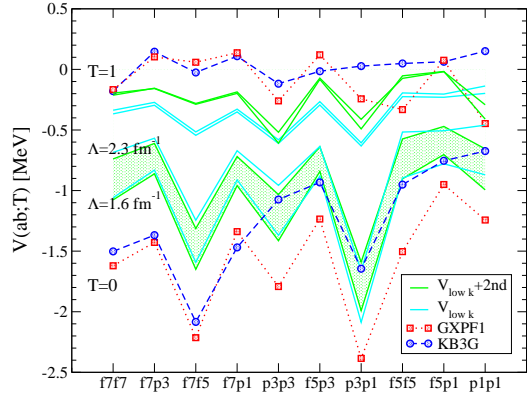}
\caption{Monopole interactions in the pf shell obtained from $\vlowk$ 
plus second-order many-body contributions for a range of cutoffs, 
compared to phenomenological matrix elements~\cite{Vlowkmono}.
\label{monopoles}\vspace*{-5mm}}
\end{figure}

For lower cutoffs, we take the corresponding 3N interactions $V_{\rm 3N}
(\lm)$ from chiral EFT by fitting the D- and E-term couplings to the 
$^3$H and $^4$He binding energies for a range of cutoffs~\cite{Vlowk3N}.
Since chiral EFT is a complete basis, this gives the 3N force up to 
truncation errors. We have found that the resulting 3N interactions
become perturbative for $\lm \lesssim 2 \fmi$, while they are 
nonperturbative for larger cutoffs, and that the size of 3N expectation
values is natural $\sim (Q/\lb)^3 \la \vlowk \ra$~\cite{Vlowk3N}. 

As for the Tjon line, cutoff variation can provide lower bounds for
theoretical uncertainties due to neglected many-body interactions
or an incomplete many-body treatment. As shown in Fig.~\ref{radii},
this is a powerful and practical tool, which is important for
extrapolations, in particular for matrix elements
needed in fundamental symmetry tests (e.g., double-beta decay
and isospin-violating corrections for super-allowed beta decay).

Coupled-cluster (CC) theory combined with rapid convergence for low-momentum 
interactions pushes the limits of accurate calculations to medium-mass
nuclei and sets new benchmarks for $^{16}$O and $^{40}$Ca~\cite{VlowkCC}. 
In Fig.~\ref{CC}, first CC results with 3N forces show that 
low-momentum 3N interactions are accurately
treated as effective 0-, 1- and 2-body terms, and that residual
3N interactions can be neglected~\cite{VlowkCC}. This is very promising
and supports the idea that phenomenological monopole shifts in shell model
interactions are due to 3N contributions~\cite{Zuker}, see also the
talk by T.~Otsuka and Ref.~\cite{SM}.
This links understanding the shell model and
the drip lines to 3N forces. First investigations in this direction
are presented in Fig.~\ref{monopoles}~\cite{Vlowkmono}. We speculate
that the $T=0$ monopoles with large cutoff dependences receive
cutoff-dependent contributions mainly from second-order NN-3N
terms (which are expected to be attractive) and that the $T=1$ monopoles
receive cutoff-independent contributions from the 3N $c_i$ terms
(which are repulsive in nuclear matter). Systematic investigations
of 3N forces and valence-shell interactions are in 
progress~\cite{Vlowkmono}. 

Finally, there is the possibility of perturbative nuclear matter with
low-momentum NN and 3N interactions~\cite{nucmatt}. As shown in 
Fig.~\ref{matter}, with second-order contributions the cutoff dependence
is very weak at low densities. We have found that 3N forces drive 
saturation~\cite{nucmatt}, and at present the uncertainties in the 
$c_i$ overwhelm other uncertainties (see the results for neutron 
matter)~\cite{neutmatt}. Lower
values for $c_3$ (as also expected from N$^3$LO 3N terms) add 
$\approx -2 \mev$ at saturation density in neutron matter, and this 
also corresponds to the value from Delta resonance saturation (which
explains the agreement with the results of Akmal et al.). These
findings will provide guidance for constructing a universal density
functional for nuclei. Towards denser matter, 4N force contributions of
$E/A \sim 1 \mev$ would not be unreasonable.

This is an exciting era: with advances on many fronts, a coherent 
effort to understand and predict nuclear systems based on effective 
field theory and renormalization group interactions, where 3N
forces are a frontier.

\vspace*{3mm}

It is a pleasure to thank the organizers for a very stimulating FM50 
conference and the collaborators who have contributed to these results:
S.~Bogner, G.~Brown, D.~Dean, B.~Friman, R.~Furnstahl, G.~Hagen, T.~Kuo, 
P.~Maris, A.~Nogga, T.~Papenbrock, R.~Perry, L.~Tolos, J.~Vary and 
A.~Zuker. This work was supported by NSERC. TRIUMF receives federal 
funding via a contribution agreement through the NRC of Canada.


\begin{thebibliography}{99}
\bibitem{pionless} P.F.~Bedaque and U.~van Kolck, \emph{Ann. Rev. Nucl. 
Part. Sci.} \textbf{52}, 339 (2002).
\bibitem{chiral} E.~Epelbaum, \emph{Prog. Part. Nucl. Phys.} \textbf{57}, 
654 (2006).
\bibitem{Vlowk} S.K.~Bogner, T.T.S.~Kuo and A.~Schwenk, \emph{Phys. Rept.}
\textbf{386}, 1 (2003).
\bibitem{dEFT} A.~Schwenk and C.J.~Pethick, \emph{Phys. Rev. Lett.} 
\textbf{95}, 160401 (2005).
\bibitem{VlowkRG} S.K.~Bogner, A.~Schwenk, T.T.S.~Kuo and G.E.~Brown,
nucl-th/0111042.
\bibitem{smooth} S.K.~Bogner, R.J.~Furnstahl, S.~Ramanan and A.~Schwenk,
\emph{Nucl. Phys. A} \textbf{784}, 79 (2007).
\bibitem{SRG1} S.K.~Bogner, R.J.~Furnstahl and R.J.~Perry, \emph{Phys. 
Rev. C} \textbf{75}, 061001(R) (2007).
\bibitem{SRG2} S.K.~Bogner et al., \emph{Phys. Rev. C} in press, 
arXiv:0801.1098.
\bibitem{nucmatt} S.K.~Bogner, A.~Schwenk, R.J.~Furnstahl and 
A.~Nogga, \emph{Nucl. Phys. A} \textbf{763}, 59 (2005).
\bibitem{Born} S.K.~Bogner, R.J.~Furnstahl, S.~Ramanan and A.~Schwenk,
\emph{Nucl. Phys. A} \textbf{773}, 203 (2006).
\bibitem{AV18} R.B.~Wiringa, V.G.J.~Stoks and R.~Schiavilla, \emph{Phys. 
Rev. C} \textbf{51}, 38 (1995).
\bibitem{EM} D.R.~Entem and R.~Machleidt, \emph{Phys. Rev. C} \textbf{68}, 
041001(R) (2003).
\bibitem{NCSM} S.K.~Bogner et al., \emph{Nucl. Phys. A} 
\textbf{801}, 21 (2008).
\bibitem{Vlowk3N} A.~Nogga, S.K.~Bogner and A.~Schwenk, \emph{Phys. Rev. C}
\textbf{70}, 061002(R) (2004).
\bibitem{Oxygen_Tjon} G.~Hagen, D.J.~Dean and A.~Schwenk, in prep.
\bibitem{Meyer} H.-O.~Meyer, TRIUMF 3N Workshop (2007), see
http://www.triumf.info/hosted/TNI/index.html.
\bibitem{Ulf} U.G.~Mei{\ss}ner, private comm. (2007).
\bibitem{Andreas} A.~Nogga, private comm. (2004).
\bibitem{VlowkCC} G.~Hagen et al., \emph{Phys. Rev. C} \textbf{76}, 034302
(2007); ibid., \textbf{76}, 044305 (2007).
\bibitem{Zuker} A.P.~Zuker, \emph{Phys. Rev. Lett.} \textbf{90}, 042502 
(2003).
\bibitem{SM} A.~Schwenk and A.P.~Zuker, \emph{Phys. Rev. C} \textbf{74},
061302(R) (2006).
\bibitem{Vlowkmono} J.D.~Holt and A.~Schwenk, in prep.
\bibitem{neutmatt} L.~Tolos, B.~Friman and A.~Schwenk, arXiv:0711.3613.
\end{thebibliography}
\end{document}